%% file: RyoHahsimotoSOIPIXproc.tex
\documentclass[12pt]{article}
\usepackage{graphicx}

\newcommand\pubdate{\today}

\def\tsukuba{Institute of Material Structure Science\\
High Energy Accelerator Research Organization, 1-1 Oho, Tukuba, Ibaraki, Japan}
\def\support{\footnote{This study was supported by JSPS KAKENHI Grant Number 25109008.}}

\def\Title#1{\begin{center} {\Large #1 } \end{center}}
\def\Author#1{\begin{center}{ \sc #1} \end{center}}
\def\Address#1{\begin{center}{ \it #1} \end{center}}

\newcommand\pubblock{\rightline{\begin{tabular}{l}
         \pubdate  \end{tabular}}}
\newenvironment{Abstract}{\begin{quotation}  }{\end{quotation}}
\newenvironment{Presented}{\begin{quotation} \begin{center} 
             PRESENTED AT\end{center}\bigskip 
      \begin{center}\begin{large}}{\end{large}\end{center} \end{quotation}}

\input econfmacros.tex

\begin{document}
\begin{titlepage}
\pubblock

\vfill
\Title{Application of SOI Area Detectors to Synchrotron Radiation X-ray Experiments}
\vfill
\Author{Ryo Hashimoto\support, Noriyuki Igarashi, Reiji Kumai and Shunji Kishimoto}
\Address{\tsukuba}
\vfill
\begin{Abstract}
Application of new detectors using Silicon-On-Insulator (SOI) technology has been started in the Photon Factory, 
KEK. This project has two purposes. The first purpose is to develop a pulse-counting-type X-ray 
detector which can be used in synchrotron soft X-ray experiments. The second one is to apply the SOI area detector 
developed by RIKEN, SOPHIAS, to X-ray diffraction and small-angle scattering experiments in Photon Factory. 
In this paper, we introduce the current status of our project. 
\end{Abstract}
\vfill
\begin{Presented}
International Workshop on SOI Pixel Detector (SOIPIX2015), \\
Tohoku University, Sendai, Japan, 3-6, June, 2015.
\end{Presented}
\vfill
\end{titlepage}
\def\thefootnote{\fnsymbol{footnote}}
\setcounter{footnote}{0}

\section{Introduction}

Structural analysis for functional materials is one of studies which are recently very interested in application of 
synchrotron radiation.
By use of hard X-ray and soft X-ray, information about its nano-structure and 3D structure including the depth 
direction can be obtained. With having higher emittance of the synchrotron radiation beam and spreading the field 
of application, higher performance is requested to a detector. 
Then, we have started developing a new pulse-counting-type area detector using Silicon-On-Insulator (SOI) 
technology\cite{SOI}, which is satisfied 
\begin{itemize}
 \item high frame rate,
 \item high spatial resolution,
 \item high sensitivity (of course for soft X-ray).
\end{itemize}

RIKEN has already developed the charge-integration-type SOI area detector, SOPHIAS\cite{SOPHIAS} which has high spatial resolution. We also have tried to use it in the structure analysis using the X-ray diffraction and small-angle X-ray 
scattering, in collaboration with RIKEN.

In this paper, performance evaluation of CPIXPTEG1, which is a Test-Element-Group (TEG) of our counting-type detector under development, is explained in next chapter. Synchrotron radiation experiments using SOPHIAS in Photon Factory, 
KEK (KEK/PF) is shown in Chapter \ref{SOPHIASexp}.

\section{Evaluation of the CPIXPTEG1}

\begin{figure}[htb]
\centering
\includegraphics[height=2.0in]{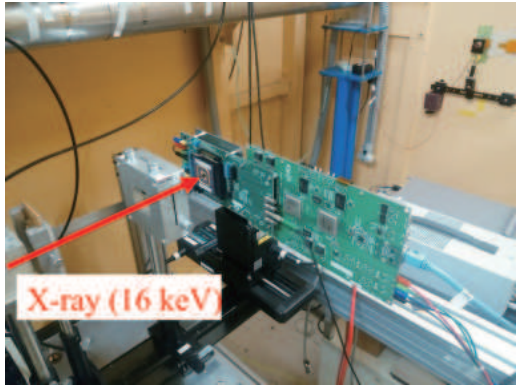}
\vspace{-0.1in}
\caption{Experimental setup of a CPIXPTEG1 evaluation.}
\label{figCPIXPTEG1setup}
\end{figure}

In developing a pulse-counting-type SOI area detector, we have started development and evaluation of a TEG called CPIXPTEG1. Its specifications are below.
\begin{itemize}
 \item p-type sensor.
 \item 32$\times$32 pixel array (pixel size is 64 $\mu$m square).
 \item 6 types of TEG and 30 types of TEG array.
 \item both single-SOI and double-SOI\cite{DSOI} model are developed.
 \item discriminator$\times$1 (lower level threshold). 
\end{itemize}
Experiments have been done at KEK/PF BL-14A by use of 16~keV X-rays. Experimental setup is shown in 
Figure.\ref{figCPIXPTEG1setup}. 
In this experiment, we aimed to check analog output signals from a preamplifier and a shaper amplifier for X-rays. 
Both one pixel in the pixel array and one of the TEG array, TEG01, have been evaluated. TEG01 can be used an isolated pixel because it has the same circuit as the pixel array and an electrode connected to the sensor. 

\begin{figure}[htb]
\centering
\includegraphics[height=2.4in]{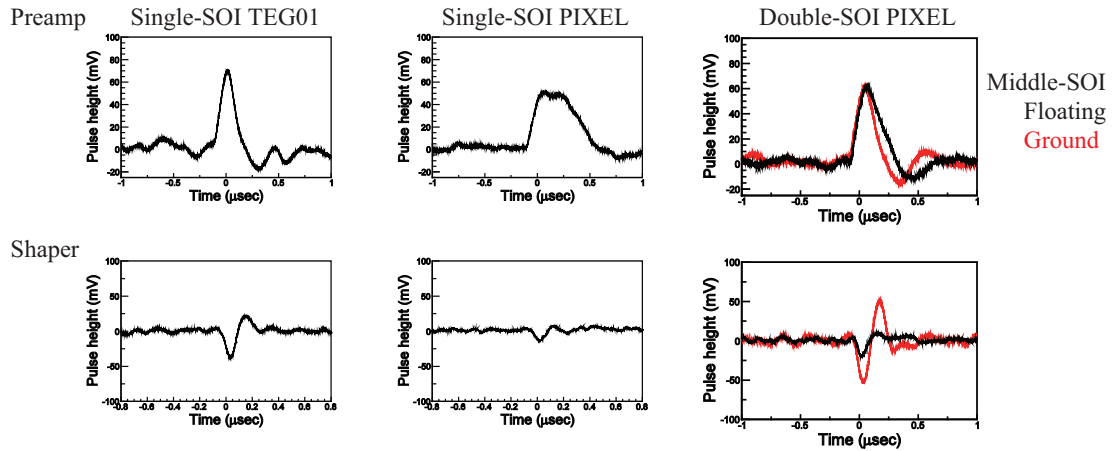}
\vspace{-0.2in}
\caption{Analog output signals of a preamplifier and a shaper amplifier. A back-bias voltage of -110~V is applied to a 
sensor. Compared with results about TEG01, a preamp signal of the pixel array was distorted and a signal from the 
shaper was smaller in a single-SOI chip. 
For a double-SOI chip, output signals when the middle-SOI layer was set to the floating level is shown by 
black line. Pulse shapes changed better when middle-SOI layer was connected to the ground level.}
\label{figanalogoutput}
\end{figure}

Analog outputs of CPIXPTEG1 are shown in Figure.\ref{figanalogoutput}. A back-bias voltage of -110~V is applied to 
a sensor in all figures. 
Compared with results about TEG01, a preamp signal of the pixel array was distorted and a signal from the shaper was smaller in a single-SOI chip. 
This reason was examined in the present work. 
For a double-SOI chip, at first we set the middle-SOI layer to the floating level (black line) 
and connected the middle-SOI layer to the ground level. Pulse shapes, especially of the shaper signal, have changed better. 

\begin{figure}[htb]
\centering
\includegraphics[height=1.8in]{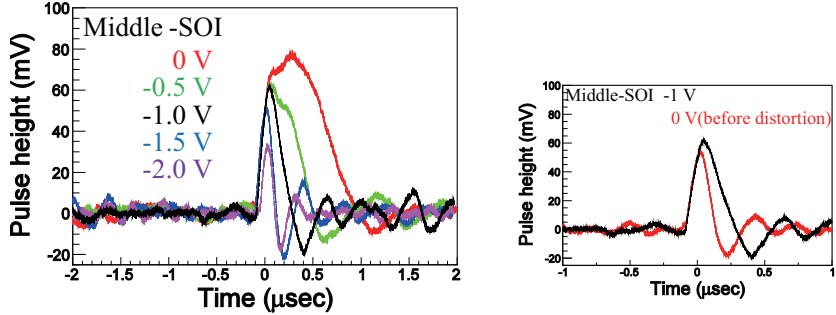}
\vspace{-0.2in}
\caption{Comparison of preamp signals of a double-SOI. Bias-voltages of the middle-SOI layer are 0~V, -0.5~V, -1.0~V, 
-1.5~V and -2.0~V in the left figure. A bias-voltage of -1.0~V was suitable, and the pulse shape recovered to the same 
quality as at the start of double-SOI evaluation (right figure)}
\label{figDSOIresult}
\end{figure}

Effect on the equipment with the double-SOI structure was also checked in this experiment. In the experiment 
for evaluating a double-SOI chip, a pulse 
shape was distorted by collected charges. After this, we applied a bias-voltage to the middle-SOI, and then the 
pulse shape was recovered (shown in Figure.\ref{figDSOIresult}). A value of -1.0~V was suitable for charge collection in 
this experiment. A pulse shape was recovered to the same quality as at the start of double-SOI evaluation (right figure). 

\section{XRD and SAXS experiments with the SOPHIAS}
\label{SOPHIASexp}

\begin{figure}[htb]
\centering
\includegraphics[height=1.4in]{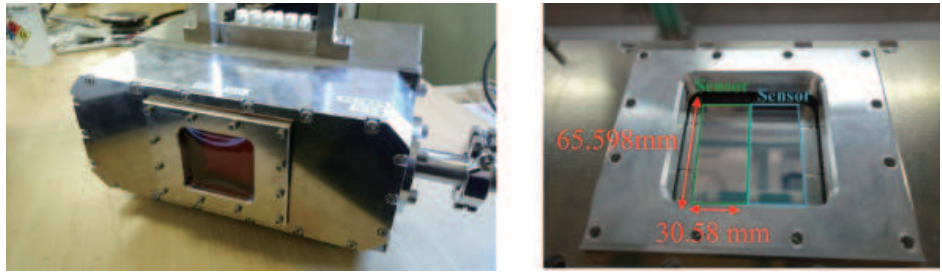}
\vspace{-0.1in}
\caption{Left side photo graph shows detector head of SOPHIAS3. Pick up view of the detection area which consists of two sensor chips is shown in right figure. The size of one chip is 65.598~mm$\times$30.58~mm.}
\label{figSOPHIAS}
\end{figure}

Because of its pixel size of 30$\times$30~$\mu$m$^2$, a charge-integration type detector SOPHIAS is a powerful tool in X-ray structural analysis. We tried application of SOPHIAS to synchrotron radiation X-ray experiments. 
Experiments were performed with SOPHIAS3 which has two sensor chips shown in Figure.\ref{figSOPHIAS}.  A sensor 
chip contains 2157$\times$891 pixels and the size of one chip is 65.598~mm$\times$30.58~mm.

\begin{figure}[htb]
\centering
\includegraphics[height=5.0in]{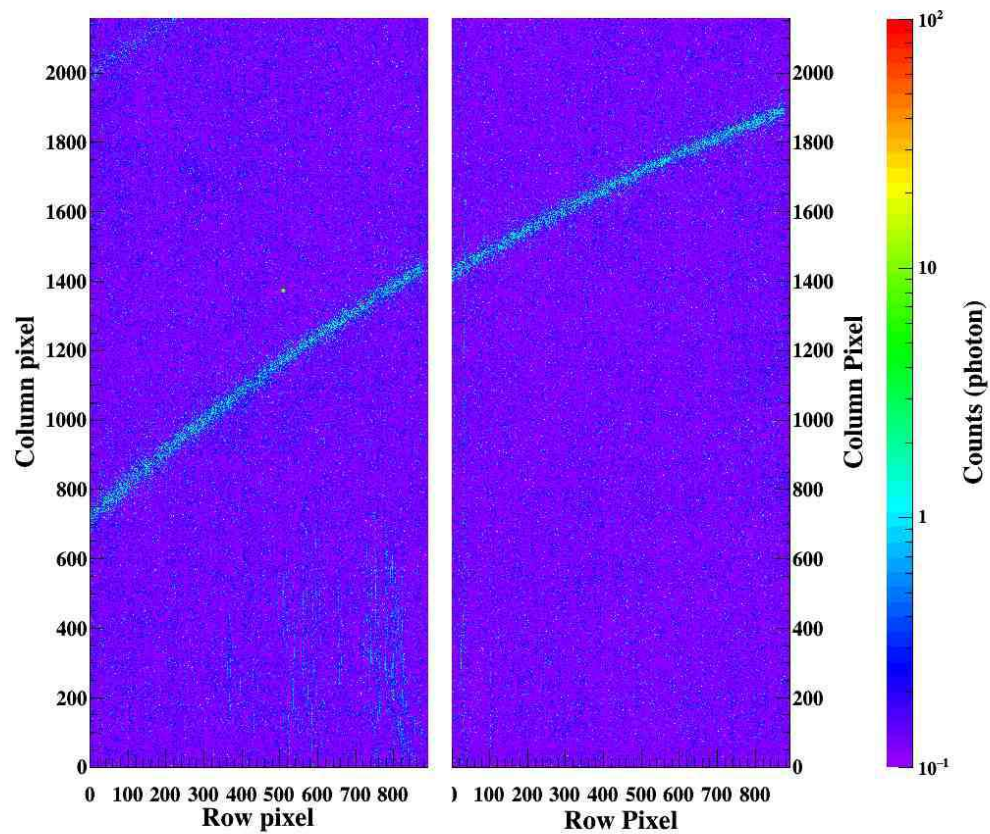}
\caption{A part of the diffraction pattern of CeO$_2$ detected by SOPHIAS3. A diffraction ring was obtained even with one photon/pixel for 12.4~keV X-ray in 300~msec exposure.}
\label{figCeO2}
\end{figure}

X-ray diffraction (XRD) experiment was done at KEK/PF BL-8A. A CeO$_2$ powder was used 
for a sample for checking the detector performance. 
CeO$_2$ is one of the standard samples for single-crystal structure analysis. A part of a 
diffraction pattern of CeO$_2$ detected by SOPHIAS3 is shown in Figure.\ref{figCeO2}. Incident X-ray energy was 
12.4~keV and exposure time was 300~msec. 
Intensity of the diffraction pattern was almost one~photon/pixel. SOPHIAS3 had enough sensitivity of one~
photon for 12.4~keV. This result enables us to perform time-resolved XRD experiments by 300~msec exposure 
in this beam line. 

\begin{figure}[htb]
\centering
\includegraphics[height=5.0in]{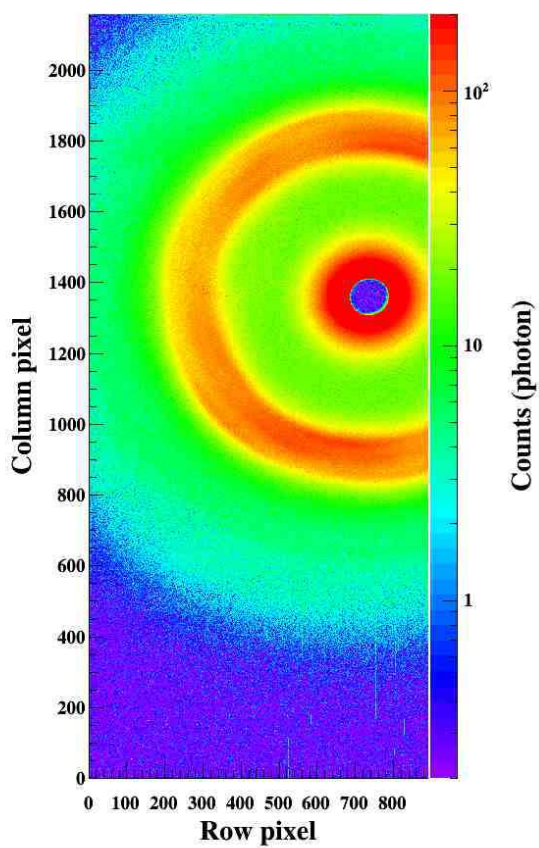}
\caption{SAXS pattern of a mixture of polystyrene-polyisoprene diblock copolymer and polyisoprene homopolymer. 
The internal fine structure was observed inside the outer ring.}
\label{figSI}
\end{figure}

Small-angle X-ray scattering (SAXS) experiments were also done at KEK/PF BL-15A2. Two samples were used in 
this experiment ; One is collagen, one of the standard samples and another is a mixture of 
polystyrene-polyisoprene dibloch copolymer and polyisoprene homopolymer (PS-PI/PI). PS-PI/PI has been studied 
for its morphologies of the microdomain structures by the SAXS method\cite{PS-PI}. It is important that a detector has 
high spatial resolution in studies of PS-PI/PI because of its complicated structure. 
A SAXS pattern of PS-PI/PI detected by SOPHIAS3 is shown in Figure.\ref{figSI} in which 7.27~keV X-rays were used. 
The inner small ring was formed by incident X-ray and a direct beam stopper. The outer ring was a scattering image of 
PS-PI/PI. Thanks to high spatial resolution, the internal fine structure caused by the PS-PI/PI structure was observed inside 
the outer ring.

\section{Summary}

We have carried out experiments for evaluating CPIXPTEG1 by use of 16~keV X-rays.
We succeeded in readout for X-ray output signals.
For the double-SOI chip, we confirmed that the middle-SOI was effective for recoverying a distorted signal caused by 
collected charges.

In parallel to evaluation for CPIXPTEG1, the test experiments using SOPHIAS3 detector have been done. 
SOPHIAS3 detected one photon of 12.4~keV X-ray in the XRD experiment. 
In the SAXS experiment, the fine structure was observed in the SAXS pattern for a PS-PI/PI 
sample. We will try more samples in the structure analysis using synchrotron science in the next experiments.

\end{document}

%% file: econfmacros.tex
\def\beq{\begin{equation}}
\def\eeq#1{\label{#1}\end{equation}}
\def\eeqn{\end{equation}}

\def\beqa{\begin{eqnarray}}
\def\eeqa#1{\label{#1}\end{eqnarray}}
\def\eeqan{\end{eqnarray}}

\let\bar=\overbar

\def\Dslash{\not{\hbox{\kern-4pt $D$}}}
\def\dslash{\not{\hbox{\kern-2pt $\del$}}}

\def\msb{{\bar{\ssstyle M \kern -1pt S}}}

